\documentclass[twocolumn]{aastex61}

\newcommand{\kms}{km\,s$^{-1}$} 

\newcommand{\hii}{\ion{H}{2}}

\newcommand{\nii}{[\ion{N}{2}]}

\newcommand{\oiii}{[\ion{O}{3}]}
\newcommand{\sii}{[\ion{S}{2}]}

\newcommand{\ha}{H$\alpha$} 
\newcommand{\hb}{H$\beta$} 
\newcommand{\flux}{erg\,s$^{-1}$\,cm$^{-2}$\,\AA$^{-1}$}

\received{July 1, 2016}
\revised{September 27, 2016}
\accepted{\today}
\submitjournal{ApJL}

\shorttitle{Optical emission line jet in NGC\,232}
\shortauthors{L\'opez--Cob\'a et al.}

\begin{document}

\title{Serendipitous discovery of an optical emission line jet in NGC\,232}

\correspondingauthor{C. L\'opez--Cob\'a}
\email{clopez@astro.unam.mx}

\author{C. L\'opez--Cob\'a}
\author{S. F. S\'anchez}
\author{I. Cruz--Gonz\'alez}
\affil{Instituto de Astronom\'ia, Universidad Nacional Aut\'onoma de  M\'exico, A.~P. 70-264, C.P. 04510, M\'exico, D.F., Mexico}

\author{L. Binette}
\affil{Instituto de Astronom\'ia, Universidad Nacional Aut\'onoma de  M\'exico, A.~P. 70-264, C.P. 04510, M\'exico, D.F., Mexico}


\author{L. Galbany}
\affil{PITT PACC, Department of Physics and Astronomy, University of Pittsburgh, Pittsburgh, PA 15260, USA}
\author{T.~Kr\"{u}hler}
\affil{Max-Planck-Institut f\"{u}r extraterrestrische Physik, Giessenbachstra\ss e, 85748 Garching, Germany}

\author{L.~F. Rodr\'iguez}
\affil{Instituto de Radioastronom\'ia y Astrof\'isica, Universidad Nacional Aut\'onoma de M\'exico, C.P. 58190, Morelia, Mexico}
\author{J.~K. Barrera--Ballesteros}
\affil{Department of Physics and Astronomy, Johns Hopkins University, 3400 N. Charles Street, Baltimore, MD 21218, USA}
\author{L. S\'anchez--Menguiano}
\affil{Departamento de F\'isica Te\'orica y del Cosmos, Universidad de Granada, Campus de Fuentenueva, 18071 Granada, Spain}
\affil{Instituto de Astrof\'isica de Andaluc\'ia (CSIC), Glorieta de la Astronom\'ia s/n, 3004, 18080 Granada, Spain}
\author{C.~J. Walcher}
\affil{Leibniz-Institut f\"{u}r Astrophysik (AIP), An der Sternwarte 16, D-14482 Potsdam, Germany}
\author{E. Aquino-Ort\'iz}
\affil{Instituto de Astronom\'ia, Universidad Nacional Aut\'onoma de  M\'exico, A.~P. 70-264, C.P. 04510, M\'exico, D.F., Mexico}
\author{J.~P. Anderson}
\affil{European Southern Observatory, Alonso de C\'ordova 3107, Vitacura, Casilla 190001, Santiago, Chile}

\begin{abstract}

We report the detection of a highly collimated linear emission-line structure in the spiral galaxy NGC\,232 through the use
of integral field spectroscopy data from the All-weather MUse Supernova Integral field Nearby Galaxies (AMUSING) survey. This jet--like feature extends radially from the nucleus and
is primarily detected in \oiii$\lambda$5007 
without clear evidence of an optical continuum counterpart. 
The length of the radial structure projected on sky reaches $\sim 3$ kpc, 
which makes NGC\,232 the second longest emission-line jet reported. The ionized gas presents extreme  \oiii/\hb\ and \nii/\ha\ line ratios, increasing along the
jet-like structure.
We { discuss three possible scenarios to explain the observed structure: (i) direct ionization of in-falling material from the intergalactic medium by the AGN; (ii) photo-ionization by an un-detected optical counter-part of the radio jet and (iii) fast shocks ionization due to the lateral expansion of the radio jet across the ISM.} Our analysis favors in-situ ionization.

\end{abstract}

\section{Introduction}

%

In active galactic nuclei (AGN), the primary source of ionizing photons responsible for exciting the extended emission-line regions (EELR) comes from the accretion disk in the nucleus. However, in the case of collimated radio jets with enhanced optical emission lines, the source of
the ionization is still an open question. It has been proposed that the ionizing continuum in this case is generated {\it in situ} \citep[e.g.,][]{Binette1993}. Two possibilities have been suggested for generating such locally produced continua: 1) by lateral fast shocks along the relativistic radio jet as it propagates in the ISM \citep[e.g.,][]{Axon1989,Fraix1992,sutherland1993}, and 2) synchrotron emission from the radio jet itself \citep[e.g.,][]{Jarvis1990}. {Nevertheless, direct ionization by the AGN cannot be excluded.}  

The EELR of Seyfert galaxies is usually studied in \oiii$\lambda5007$, a feature that shows a wide of shapes. These are typically broadly
conical or clumpy and radially elongated, ranging in lengths from hundreds to thousands of parsecs. EELRs are frequently, but not always, aligned with radio jets \citep[e.g.,][]{Capetti1995b,Husemann2008} when such a jet is present. 
Generally, the emission line gas associated with the jets is observed as outflowing material from the core of galaxies, but in some cases it can be observed as inflowing gas \citep[e.g.,][]{Jarvis1990}.
In a small number of cases, such as 3C\,120, the host galaxy, AGN and the jet have been observed in many different bands from radio to X--rays, including integral field spectroscopy \citep[e.g.,][]{Axon1989,Hjorth1995,Sanchez2004,Begona2005}, providing a better understanding of the physical processes involved. 

In this Letter we present the detection of a highly collimated  jet-like structure in NGC\,232 observed in optical emission-lines. 
The possible driving sources of the observed structure and its correspondence to a radio emission are discussed. 

\section{Data}
NGC\,232 is an Sa galaxy located at z = 0.022, at a distance of 91.3 Mpc \citep[e.g.,][]{Amanullah}, and is part of a group of galaxies that also includes NGC\,235. It has been classified
as a luminous infrared galaxy \citep[e.g.,][]{Sanders} and as a radio{--quiet } source \citep[e.g.,][]{Condon1998,Hill2001}. { The luminosity at 1.4 GHz is $L_{1.4} = 5.96 \times 10 ^{22}$ W Hz$^{-1}$ \citep[e.g.,][]{Corbett2002}.
} 

NGC\,232 was observed on 2015 June 26th during the first semester of observations (095.D-0091; PI Anderson) of the All-weather MUse
Supernova Integral field Nearby Galaxies \citep[AMUSING,][]{Galbany2016} survey with the Multi Unit Spectroscopic 
Explorer \citep[MUSE,][]{Bacon2010} on the ESO’s Very Large Telescope UT4 (Yepun). 
%
%
MUSE is an integral-field spectrograph (IFS) with a large field-of-view (FoV) of 1 arcmin$^2$ and a high spatial sampling of 0.2$\arcsec$ spaxel$^{-1}$. It covers the
wavelength range between 4750 and 9300 \AA, with a constant spectral sampling of 1.25 \AA~ and a spectral resolution of FWHM $\sim 2.6$ \AA.
A total integration time of $\sim 2000$~s was performed on source. Data reduction is described in \citet{Galbany2016} \& \citet{Kruehler2017}.
The final dataset comprises one cube of $\sim 100,000$ individual spectra with a spatial resolution determined by the average seeing during the observation $\sim 0$\farcs67 ($\sim 300$ pc at the distance of NGC\,232, {using a standard cosmology of H$_o=$70 \kms Mpc$^{-1}$, $\Omega_m=$0.27, $\Omega_\Lambda=$0.73}).
%
In addition we use radio maps extracted from the Very Large Array (VLA) archives, observed at a frequency of 1.4 GHz (VLA AM384 project). Its synthesized beam is $3 \farcs 22 \times 1\farcs 66$, with a PA\,=\,-19.7\,$^\circ$. Thus, the VLA image has a coarse spatial resolution compared to the IFS data. 
%

\begin{figure*}
  \centering
     \includegraphics[trim=0cm 0cm 0cm 0cm,clip=true,width=\textwidth,height=!]{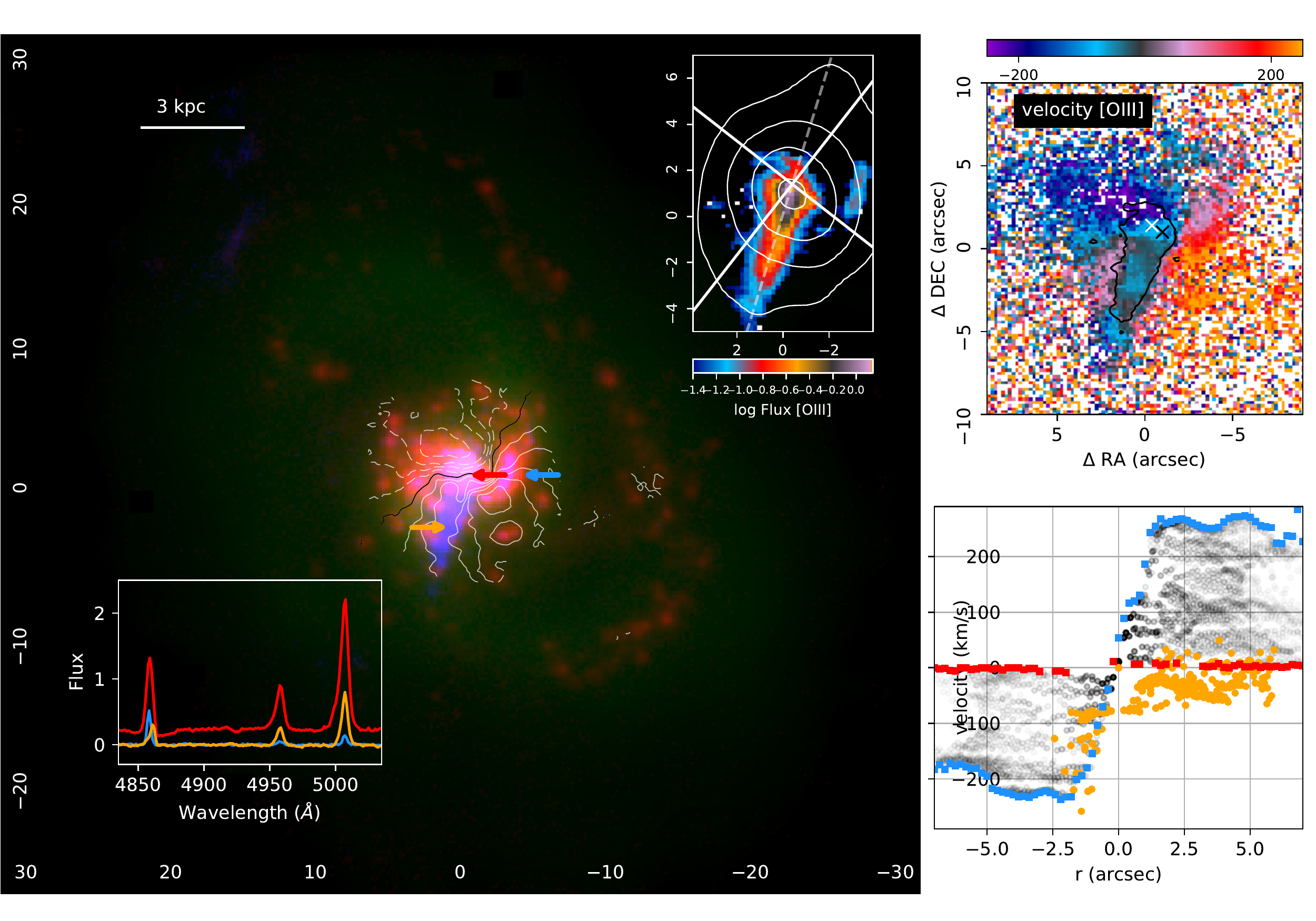}%
  \caption{\textit{Main panel}: RGB  color image of NGC\,232: \ha~intensity in red, V-band intensity in green and \oiii$\lambda5007$ intensity in blue. White contours trace the \ha~velocity map in steps of 50 \kms, with the solid contours displaying receding velocities and the dashed
countours the approaching velocities. The black solid line indicates the location of zero velocity. The bottom-left inset shows the co-added spectra (within a resolution element of $\sim 0$\farcs67) for three different positions located at (i) the nucleus (red), (ii) the disk (blue), and (iii) the elongated emission-line structure (orange). Each location is marked with an arrow in the main panel.
These spectra cover the wavelength range of \hb~to \oiii$\lambda5007$~ with the flux shown in units of 10$^{-16}$ \flux. The upper-right inset shows the \oiii~intensity map in logarithmic scale using colors and zoomed in at the location of the jet structure. On-top of that image is superimposed the contour map of the 1.4\,GHz radio emission obtained by the VLA, with the contour levels being 0.35, 2.1, 6.7 and 15.4 mJy/beam intensities, respectively. The white straight lines indicate the location of the semi-major (North-East to South-West) and semi-minor (North-West to South-East) axis of the galaxy. The  white dashed-line shows the mean position angle $\mathrm{PA_{[OIII]}}$ of the  \oiii\ structure. \textit{Upper right panel} shows the velocity field of \oiii$\lambda$5007 color-coded in \kms. The spatial extension of the jet-like structure is shown with a black solid contour line. Its location is defined by  the position where the \oiii~ intensity reaches $\sim 0.04 \times 10^{-16}$ \flux. The white and 
black crosses show the photometric (V-band) and
kinematic centers, respectively. \textit{Bottom right panel} shows the { 2D} position-velocity diagram of NGC\,232 relative to its kinematic center. The semi-transparent grey squares represent the \ha~velocity, with the
blue (red) squares tracing the maximum (minimum) 
envelope of these velocities. The blue squares trace the rotation curve of the galaxy. Orange squares represent the \oiii\ position-velocity diagram for the jet-like structure. It comprises those spaxels located within the black contour shown in the upper-right panel.}
  \label{rgb} 
\end{figure*}
\section{Analysis}
The data were analyzed using the {\sc Pipe3D} pipeline \citep[e.g.,][]{PIPE3D}, a code created to determine the stellar content and ionized gas properties of IFS data.   
In summary, the continuum is modeled as a combination of 
synthetic stellar populations
of different ages and metallicities. Once the best stellar model for each
spaxel has been determined, it is subtracted from the original data cube to obtain a pure emission-line gas cube.
Each emission-line profile is then modeled using a single Gaussian function to determine the corresponding fluxes and errors, together with its velocity and velocity dispersion. 
The result of the fitting procedure is a set of bi--dimensional maps comprising
the spatial distribution of the flux intensities for each analyzed emission line 
including \ha, the \nii$\lambda\lambda$6548,6584 doublet, \hb,
\oiii$\lambda\lambda$4959,5007 and other weak emission lines. 

\section{Ionized gas and kinematics}
In Fig. \ref{rgb} we present a color-coded composite image of NGC\,232. The high spectral and spatial 
resolution of MUSE allows us to clearly identify the spiral arms together with the clumpy structures associated with \hii~regions.
Noticeably, an
extended structure in \oiii$\lambda 5007$ emission (hereafter \oiii) is present (displayed in blue). This shows a collimated structure that resembles that found in radio galaxies. This jet--like structure is stronger in \oiii, but it is also detected in
\ha,~\nii, \sii~ and at a lower signal to noise (S$/$N) in \hb. The
elongated emission structure is characterized by a  $\mathrm{PA_{[OIII]}}$ of 163$^\circ$, which differs by  $\sim20^\circ$
from the semi-minor axis of NGC\,232. Such a misalignment
with the semi-minor axis indicates that the jet-structure is not perpendicular to the galaxy disk, which is not uncommon in these kind of objects \citep[e.g.,][]{Kinney2000}.
The bottom left inset in the main panel of Fig. \ref{rgb} illustrates the \hb --\oiii\ spectral range at three different locations in the galaxy: the nucleus, the disk and the elongated emission-line region. 
They highlight the fact that the relative emission-line intensities
clearly vary with position.
In the disk, the \hb~flux is higher than \oiii\, as expected for star forming regions, while \oiii\ dominates in both the nucleus and the elongated structure. 
In the nucleus, the strong  \oiii\ line is compatible with that of an AGN. The presence of a strong radio-source centered on the nucleus favors  the presence of an active nucleus. The 
VLA radio contours are elongated along the same direction as  \oiii\  (top right inset of main panel).
We also note that within the elongated emission-line structure, the line profiles are clearly asymmetric, which is indicative of
multiple kinematic components.
\begin{figure*}
  \centering
     \includegraphics[width=\textwidth,height=!]{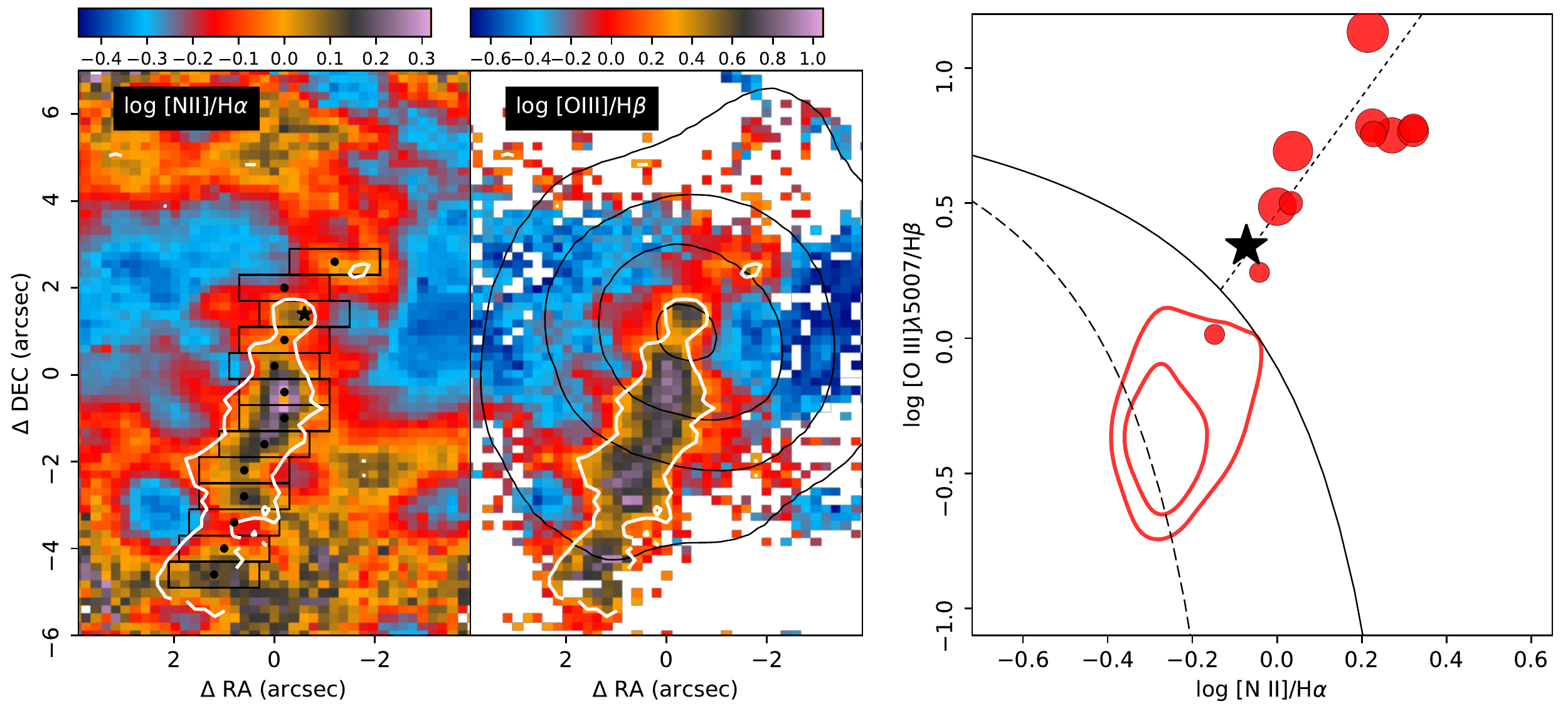}%
  \caption{Color-maps showing the spatial distribution of the line ratios   \nii/\ha~(left panel) and \oiii/\hb~(following panel)  for all spaxels with a S/N $>2$ in the intensity of each involved emission line. In both panels, the
white contour refers to the line ratio threshold of log(\oiii/\hb) = 0.3, which is used to define our jet-like structure. Boxes of sizes $1 \farcs 8 \times 0 \farcs 6$ are drawn in the leftmost-panel along the jet axis. The position of the box centers (black dots) are defined by their respective peak in \oiii/\hb\ ratio (see next panel) within each box. The photometric center is shown with a black star. The black contours in the \oiii/\hb\ map represent the same radio contours as in the top-right inset of Fig. \ref{rgb}. The rightmost panel shows the classical BPT diagram together with the \citet{kauffmann} (black dashed-line) and \citet{Kewley2001} (continuous black line) demarcation curves. The dotted line represents the LINER/Seyfert transition according to \citet{Kewley2006} (with LINERs typically below this curve). In this BPT panel, the red contours represent the location of all line ratios that lie outside the white contour on both the \nii/\ha\ and \oiii/\hb\ panels. The outer (inner) red contour 
encloses 68 (34) percent of the spaxels, respectively. The red filled-circles in the BPT diagram represent the location for the co-added fluxes within each black box (see leftmost panel). The black filled-star represents the values corresponding to the box at the nucleus location. The size of each filled marker represents its projected radial distance with respect to the photometric center, being the largest ones the farthest from the nucleus. The closest (farthest) is at 1\farcs3 (6\farcs2).}
  \label{bpt} 
\end{figure*}

The typical shape of a rotating disk is clearly apparent in the \ha\ velocity contours shown by the white lines in the main panel (the zero velocity is aligned with the semi-minor axis). We should note that the photometric and kinematic centroid show an offset of $\sim 1\arcsec$, as is usually found in galaxies with an AGN or a disturbed
morphology \citep[e.g.,][]{Ballesteros2014}. 

The \oiii\ line velocity field shown in the rightmost top panel of Fig. \ref{rgb} reveals the kinematics of a different gas component, mostly that of the jet--like structure. This becomes obvious from inspection of the  rotation curve (Fig. \ref{rgb}, bottom rightmost panel), in which the kinematics of the jet--like structure appears blue--shifted with respect to the overall disk kinematics. { In fact, along the entire jet--like structure asymmetric emission lines profiles are observed. A double Gaussian model fitting of these emission line profiles reveals two clearly distinguished components: a narrow one, redshifted, with a $\sigma \sim 45$ \kms\ width (similar to the one found along the disk at the same galactocentric distances) and a broader one, blueshifted ($\sim 50$ \kms), with a $\sigma \sim 130$ \kms, only present at the location of the jet--like structure. Both reported velocity dispersions are corrected for instrumental resolution. }
%
 %
%
%
\section{Line emission along the jet--like structure}

%
In Fig. \ref{bpt}, we show the spatially resolved \nii/\ha~ and \oiii/\hb~ line ratio maps  for a region of $8\arcsec\times13\arcsec$
around the nucleus. The \oiii/\hb~map reveals the presence of a collimated line emission structure which could be associated to the path followed by a radio jet 
as it propagates radially from the nucleus. The projected width of the jet--like structure is $\sim 2\arcsec$  ($\sim$ 0.8 kpc), which means that it is transversally resolved. The full length of the jet--like structure is somewhat uncertain as it presents two discontinuities in  ionization level (as represented by the \oiii/\hb\ ratio): one near the nucleus and the second one near the outer edge. 
The length of the main jet--like structure is $\sim$ 4\farcs5 long ($\sim$ 2 kpc). If we include the outer segment of $\sim$ 1\farcs5 long (0.6 kpc), the length becomes 2.6 kpc. On the other hand, if the \oiii\ emission  from the nucleus was also part of the the jet--like structure, then the total extent of the jet--like structure reaches $\sim$ 3 kpc long. This size is actually a lower limit since it corresponds to the projected value and we do not know the orientation angle. Comparing the total size with other optical emission-line jets \citep[Mrk 3, 250 pc; M\,87, 1 kpc; 3C\,120, 4.5 kpc.][see their Fig. 6]{Capetti1995,Jarvis1990,Begona2005}, makes NGC\,232 as the galaxy hosting the second longest optical emission-line jet, so far detected.

Outside the jet--like structure, the \nii/\ha\ and \oiii/\hb\ line ratios are consistent with photoionization by young stars and a high metallicity gas \citep[e.g.,][]{sanchez2015a}. 
The location of the strongest ionization level, as traced by the \oiii/\hb~line ratio, presents wiggles along the radial jet-like emission.
Interestingly, we observe a transversal decrease in the excitation level away from the inner peak values. This feature is less evident in the \nii/\ha\ panel.

Within the extended jet of 3C\,120 the areas with higher ionization are associated with denser ionized gas \citep[e.g.,][]{Begona2005}. Unfortunately, we are unable to
estimate the electron density along the jet of NGC\,232, { due to the effect of a telluric absorption on top of the \sii$\lambda\lambda$6717,6731 lines}. Interestingly, the excitation level of the nuclear region gas (Fig. \ref{rgb}) is lower than the higher values found along the jet--like structure. 

The \oiii/\hb\ ratio is visible on both sides of the nucleus although at a much lower ionization level in the direction of the counter-jet. Interestingly, the  radio contours are somewhat elongated along the jet-like structure, but more so along the counter-jet position. This suggests that there is evidence of not only an optical emission-line jet, but also of a much fainter and smaller counter-jet which may be obscured by the disk itself.
 
The location of the highest ionization, as traced by the \oiii/\hb\ ratio, presents wiggles in its peak values.
In galaxies hosting radio jets, such wiggles are seen in their optical continuum emission \citep[e.g.,][]{Capetti1995}. 
Outflows driven by AGN or star formation similarly show a radial increase in both line ratios (\nii/\ha\ and \oiii/\hb) \citep[e.g.,][]{Cecil2001,carlos2016}. We emphasize, however, that in such cases
the ionization maps typically reveal a broadly conical or filamentary  emission-line structure, unlike NGC\,232 which clearly shows a collimated structure. In either case, the increase in the above ratios 
is compatible with either shock (or fast shock)  ionization or photoionization by a power-law ionizing source.
%
The interaction of radio jets with the ISM or the halo gas can induce jet--driven shocks at its outskirts \citep[see for example,][]{Clark1996,Axon2001}. The head-on shock morphology in this case would result in a bow-shock or a shell.  
However, such morphology is not observed around our collimated emission structure. At best, as discussed below, lateral fast shocks could be occurring in the jet transversal walls and generate an {\it in situ} continuum \citep[e.g.,][]{sutherland1993,Capetti1995,Sanchez2004}.

In Fig. \ref{bpt}, we also show the distribution of the ionized regions across the  classical BPT diagram \citep[e.g.,][]{BPT1981} together with the usual demarcation curves from \citet{Kewley2001} and \citet{kauffmann} which attempt to separate ionization by star formation from that by AGN ionization. The 
intermediate region between both curves
is frequently interpreted as a combination of different ionization
sources. The majority of the spaxels (external red contour, $68\%$) lying in the disk fall within this mixing region.
This could be due to the presence of a hard 
ionization source, like an AGN, altering the line ratios from star-forming regions, a process which is evident within the disk. A set of co-added spectra were extracted for different rectangular apertures of $1\farcs8 \times 0\farcs6$ along the jet structure (as defined in Fig. \ref{bpt}). Their location within the BPT diagram is represented by red filled-circles. Surprisingly, the ionization level at the nucleus is not the highest found. In fact, the further away from the nucleus the boxes lie, the higher their \oiii/\hb\ ratio tends to become.   
The most common behavior in EELRs ionized by an AGN is the opposite \citep[e.g.,][]{Husemann2014}.



\section{Discussion and Conclusions}


We report the detection of a tightly collimated structure of high ionized gas in NGC\,232 of $\sim$3 kpc in length, aligned with the radio emission. The nature of this jet-like structure is uncertain. Three possible scenarios could explain the results: (i) ionization by the AGN of in-falling gas from the intergalactic medium \citep[e.g., ][]{husemann2011}; (ii) in-situ photo-ionization by an undetected radio jet that extends from the radio frequencies to the optical-UV domain \citep[e.g, M87, 3C\, 31, 3C\, 66B, 3C\, 273][]{Butcher1980,Miley1981,1981ESASP.162..145T,1983ASSL..103.....F,lalievre1984}; and (iii) in-situ fast shocks induced by the lateral expansion of a radio jet across the interstellar medium  that surrounds the radio jet \citep[e.g.][]{Axon1989,Perlman1999,Sanchez2004}.

{The observed jet-like structure makes direct AGN ionization unlikely. 
However, this could be possible in the presence of highly collimated in-falling pristine gas, that indeed  has been observed in a few AGN \citep[e.g., ][]{husemann2011}. Gas could be in-falling either (a) at the far side of the disk structure, or (b) in front. The first case is unlikely due to obscuration by the dust, {that would make it undetectable or distort its apparent shape due to the differential extinction across the disk.} In the second case, it is expected that the in-falling gas would be redshifted with respect to the systemic velocity, contrary to what is observed (Fig. \ref{rgb} and \S  4). Thus, in-falling gas  onto the AGN is unlikely,  irrespectively of the ionizing source.} 

{In order to explore the possibility of photoionization by the radio jet, we extrapolate 
the observed radio fluxes to UV frequencies \citep[e.g.,][]{Griffith1994,Condon1998,Corbett2002,Schmitt2006}. { Due to the coarse resolution of the radio maps this flux may be partially originated by a synchrotron jet}. { A power--law} slope of $\alpha = -0.8$ is derived by fitting the multi-frequency radio data. { Then we integrate the spectra beyond the ionization limit (i.e., 13.6 eV), and estimate that the ionizing flux is $\sim 7.2\times10^{50}$ photons s$^{-1}$. } The Lyman continuum photons needed to produce the observed \ha\ flux along the jet--like structure is  $\sim 3.2 \times 10^{52}$ photons s$^{-1}$ \citep{osterbrock}. The miss-match between both quantities would be even larger { considering that the covering and filling factors would be always lower than unity} \citep[e.g.,][]{Heckman1993}. Thus, {\it in-situ } ionization by the synchrotron radiation of the radio jet is { also} unlikely.  

The fact that the synchrotron radiation of the radio jet is not observed in the optical regime \citep[contrary to the case of M\,87, e.g.,][]{Jarvis1990} is due to its expected surface brightness. The $3\sigma$ detection limit of our MUSE data along the jet--like structure in the $g$--band is $g_\mathrm{AB} = 19.1$ mag arcsec$^{-2}$. The extrapolation of the radio emission to the optical regime, using $\alpha = -0.8$, 
indicates that the expected surface brightness would be $g_\mathrm{AB} = 23.1$ mag arcsec$^{-2}$. Thus, it would remain undetectable in our current data. 

{Jet/cloud interactions may produce structures similar to the ones reported here \citep[e.g.,][]{best1997,clark1998,tilak2005}, although {somehow } less collimated. {Sometimes, } they produce bow-shocks or shell-like structures \cite[e.g.][]{Clark1996,Ferruit1999}. Therefore, {\it in-situ} ionization due to fast shocks associated with lateral expansion of the radio jet is a more plausible explanation to the observed structure \citep[e.g.][]{Axon1989}. Furthermore, the presence of a broad emission-line component associated with the jet-like structure supports this scenario. It is clear that higher resolution radio data and deeper optical and emission-line images will be required to improve our understanding of the various physical processes that generate the reported emission-line jet in NGC\,232.}

\acknowledgments

We thank the referee for his/her constructive comments, which have improved the manuscript presentation. C.L.C. and S.F.S. thank PAPIIT-DGAPA-IA101217 (UNAM) project. I.C.G. acknowledges support from DGAPA-UNAM grant IN113417. C.L.C. acknowledge CONACYT (Mexico) for a Ph.~D. scholarship. T.~K. acknowledges support through the S. Kovalevskaja Award to P.~Schady from the A. von Humboldt Foundation of Germany.



\bibliographystyle{aasjournal}
\bibliography{ref} 

\begin{thebibliography}{}
\expandafter\ifx\csname natexlab\endcsname\relax\def\natexlab#1{#1}\fi
\providecommand{\url}[1]{\href{#1}{#1}}

\bibitem[{{Amanullah} {et~al.}(2010){Amanullah}, {Lidman}, {Rubin}, {Aldering},
  {Astier}, {Barbary}, {Burns}, {Conley}, {Dawson}, {Deustua}, {Doi}, {Fabbro},
  {Faccioli}, {Fakhouri}, {Folatelli}, {Fruchter}, {Furusawa}, {Garavini},
  {Goldhaber}, {Goobar}, {Groom}, {Hook}, {Howell}, {Kashikawa}, {Kim}, {Knop},
  {Kowalski}, {Linder}, {Meyers}, {Morokuma}, {Nobili}, {Nordin}, {Nugent},
  {{\"O}stman}, {Pain}, {Panagia}, {Perlmutter}, {Raux}, {Ruiz-Lapuente},
  {Spadafora}, {Strovink}, {Suzuki}, {Wang}, {Wood-Vasey}, {Yasuda}, \&
  {Supernova Cosmology Project}}]{Amanullah}
{Amanullah}, R., {Lidman}, C., {Rubin}, D., {et~al.} 2010, \apj, 716, 712

\bibitem[{{Axon} {et~al.}(1989){Axon}, {Pedlar}, {Unger}, {Meurs}, \&
  {Whittle}}]{Axon1989}
{Axon}, D.~J., {Pedlar}, A., {Unger}, S.~W., {Meurs}, E.~J.~A., \& {Whittle},
  D.~M. 1989, \nat, 341, 631

\bibitem[{{Bacon} {et~al.}(2010){Bacon}, {Accardo}, {Adjali}, {Anwand},
  {Bauer}, {Biswas}, {Blaizot}, {Boudon}, {Brau-Nogue}, {Brinchmann},
  {Caillier}, {Capoani}, {Carollo}, {Contini}, {Couderc}, {Daguis{\'e}},
  {Deiries}, {Delabre}, {Dreizler}, {Dubois}, {Dupieux}, {Dupuy}, {Emsellem},
  {Fechner}, {Fleischmann}, {Fran{\c c}ois}, {Gallou}, {Gharsa}, {Glindemann},
  {Gojak}, {Guiderdoni}, {Hansali}, {Hahn}, {Jarno}, {Kelz}, {Koehler},
  {Kosmalski}, {Laurent}, {Le Floch}, {Lilly}, {Lizon}, {Loupias}, {Manescau},
  {Monstein}, {Nicklas}, {Olaya}, {Pares}, {Pasquini}, {P{\'e}contal-Rousset},
  {Pell{\'o}}, {Petit}, {Popow}, {Reiss}, {Remillieux}, {Renault}, {Roth},
  {Rupprecht}, {Serre}, {Schaye}, {Soucail}, {Steinmetz}, {Streicher}, {Stuik},
  {Valentin}, {Vernet}, {Weilbacher}, {Wisotzki}, \& {Yerle}}]{Bacon2010}
{Bacon}, R., {Accardo}, M., {Adjali}, L., {et~al.} 2010, in \procspie, Vol.
  7735, Ground-based and Airborne Instrumentation for Astronomy III, 773508

\bibitem[{{Baldwin} {et~al.}(1981){Baldwin}, {Phillips}, \&
  {Terlevich}}]{BPT1981}
{Baldwin}, J.~A., {Phillips}, M.~M., \& {Terlevich}, R. 1981, \pasp, 93, 5

\bibitem[{{Barrera-Ballesteros} {et~al.}(2014){Barrera-Ballesteros},
  {Falc{\'o}n-Barroso}, {Garc{\'{\i}}a-Lorenzo}, {van de Ven}, {Aguerri},
  {Mendez-Abreu}, {Spekkens}, {Lyubenova}, {S{\'a}nchez}, {Husemann}, {Mast},
  {Garc{\'{\i}}a-Benito}, {Iglesias-Paramo}, {Del Olmo}, {M{\'a}rquez},
  {Masegosa}, {Kehrig}, {Marino}, {Verdes-Montenegro}, {Ziegler}, {McIntosh},
  {Bland-Hawthorn}, {Walcher}, \& {Califa Collaboration}}]{Ballesteros2014}
{Barrera-Ballesteros}, J.~K., {Falc{\'o}n-Barroso}, J.,
  {Garc{\'{\i}}a-Lorenzo}, B., {et~al.} 2014, \aap, 568, A70

\bibitem[{{Best} {et~al.}(1997){Best}, {Longair}, \& {Roettgering}}]{best1997}
{Best}, P.~N., {Longair}, M.~S., \& {Roettgering}, H.~J.~A. 1997, \mnras, 292,
  758

\bibitem[{{Binette} {et~al.}(1993){Binette}, {Fosbury}, \&
  {Parker}}]{Binette1993}
{Binette}, L., {Fosbury}, R.~A., \& {Parker}, D. 1993, \pasp, 105, 1150

\bibitem[{{Butcher} {et~al.}(1980){Butcher}, {van Breugel}, \&
  {Miley}}]{Butcher1980}
{Butcher}, H.~R., {van Breugel}, W., \& {Miley}, G.~K. 1980, \apj, 235, 749

\bibitem[{{Capetti} {et~al.}(1995{\natexlab{a}}){Capetti}, {Axon}, {Kukula},
  {Macchetto}, {Pedlar}, {Sparks}, \& {Boksenberg}}]{Capetti1995}
{Capetti}, A., {Axon}, D.~J., {Kukula}, M., {et~al.} 1995{\natexlab{a}}, \apjl,
  454, L85

\bibitem[{{Capetti} {et~al.}(1995{\natexlab{b}}){Capetti}, {Macchetto}, {Axon},
  {Sparks}, \& {Boksenberg}}]{Capetti1995b}
{Capetti}, A., {Macchetto}, F., {Axon}, D.~J., {Sparks}, W.~B., \&
  {Boksenberg}, A. 1995{\natexlab{b}}, \apj, 448, 600

\bibitem[{{Cecil} {et~al.}(2001){Cecil}, {Bland-Hawthorn}, {Veilleux}, \&
  {Filippenko}}]{Cecil2001}
{Cecil}, G., {Bland-Hawthorn}, J., {Veilleux}, S., \& {Filippenko}, A.~V. 2001,
  \apj, 555, 338

\bibitem[{{Clark} \& {Tadhunter}(1996)}]{Clark1996}
{Clark}, N., \& {Tadhunter}, C. 1996, {Jet-Cloud interactions in powerful radio
  galaxies}, ed. C.~L. {Carilli} \& D.~E. {Harris}, 15

\bibitem[{{Clark} {et~al.}(1998){Clark}, {Axon}, {Tadhunter}, {Robinson}, \&
  {O'Brien}}]{clark1998}
{Clark}, N.~E., {Axon}, D.~J., {Tadhunter}, C.~N., {Robinson}, A., \&
  {O'Brien}, P. 1998, \apj, 494, 546

\bibitem[{{Condon} {et~al.}(1998){Condon}, {Cotton}, {Greisen}, {Yin},
  {Perley}, {Taylor}, \& {Broderick}}]{Condon1998}
{Condon}, J.~J., {Cotton}, W.~D., {Greisen}, E.~W., {et~al.} 1998, \aj, 115,
  1693

\bibitem[{{Corbett} {et~al.}(2002){Corbett}, {Norris}, {Heisler}, {Dopita},
  {Appleton}, {Struck}, {Murphy}, {Marston}, {Charmandaris}, {Kewley}, \&
  {Zezas}}]{Corbett2002}
{Corbett}, E.~A., {Norris}, R.~P., {Heisler}, C.~A., {et~al.} 2002, \apj, 564,
  650

\bibitem[{{Ferrari} \& {Pacholczyk}(1983)}]{1983ASSL..103.....F}
{Ferrari}, A., \& {Pacholczyk}, A.~G., eds. 1983, Astrophysics and Space
  Science Library, Vol. 103, {Astrophysical jets; Proceedings of the
  International Workshop, Turin, Italy, October 7-9, 1982}

\bibitem[{{Ferruit} {et~al.}(1999){Ferruit}, {Wilson}, {Falcke}, {Simpson},
  {P{\'e}contal}, \& {Durret}}]{Ferruit1999}
{Ferruit}, P., {Wilson}, A.~S., {Falcke}, H., {et~al.} 1999, \mnras, 309, 1

\bibitem[{{Fraix-Burnet}(1992)}]{Fraix1992}
{Fraix-Burnet}, D. 1992, \aap, 259, 445

\bibitem[{{Galbany} {et~al.}(2016){Galbany}, {Anderson}, {Rosales-Ortega},
  {Kuncarayakti}, {Kr{\"u}hler}, {S{\'a}nchez}, {Falc{\'o}n-Barroso},
  {P{\'e}rez}, {Maureira}, {Hamuy}, {Gonz{\'a}lez-Gait{\'a}n}, {F{\"o}rster},
  \& {Moral}}]{Galbany2016}
{Galbany}, L., {Anderson}, J.~P., {Rosales-Ortega}, F.~F., {et~al.} 2016,
  \mnras, 455, 4087

\bibitem[{{Garc{\'{\i}}a-Lorenzo} {et~al.}(2005){Garc{\'{\i}}a-Lorenzo},
  {S{\'a}nchez}, {Mediavilla}, {Gonz{\'a}lez-Serrano}, \&
  {Christensen}}]{Begona2005}
{Garc{\'{\i}}a-Lorenzo}, B., {S{\'a}nchez}, S.~F., {Mediavilla}, E.,
  {Gonz{\'a}lez-Serrano}, J.~I., \& {Christensen}, L. 2005, \apj, 621, 146

\bibitem[{{Griffith} {et~al.}(1994){Griffith}, {Wright}, {Burke}, \&
  {Ekers}}]{Griffith1994}
{Griffith}, M.~R., {Wright}, A.~E., {Burke}, B.~F., \& {Ekers}, R.~D. 1994,
  \apjs, 90, 179

\bibitem[{{Heckman} {et~al.}(1993){Heckman}, {Lehnert}, \&
  {Armus}}]{Heckman1993}
{Heckman}, T.~M., {Lehnert}, M.~D., \& {Armus}, L. 1993, in Astrophysics and
  Space Science Library, Vol. 188, The Environment and Evolution of Galaxies,
  ed. J.~M. {Shull} \& H.~A. {Thronson}, 455

\bibitem[{{Hill} {et~al.}(2001){Hill}, {Heisler}, {Norris}, {Reynolds}, \&
  {Hunstead}}]{Hill2001}
{Hill}, T.~L., {Heisler}, C.~A., {Norris}, R.~P., {Reynolds}, J.~E., \&
  {Hunstead}, R.~W. 2001, \aj, 121, 128

\bibitem[{{Hjorth} {et~al.}(1995){Hjorth}, {Vestergaard}, {Sorensen}, \&
  {Grundahl}}]{Hjorth1995}
{Hjorth}, J., {Vestergaard}, M., {Sorensen}, A.~N., \& {Grundahl}, F. 1995,
  \apjl, 452, L17

\bibitem[{{Husemann} {et~al.}(2014){Husemann}, {Jahnke}, {S{\'a}nchez},
  {Wisotzki}, {Nugroho}, {Kupko}, \& {Schramm}}]{Husemann2014}
{Husemann}, B., {Jahnke}, K., {S{\'a}nchez}, S.~F., {et~al.} 2014, \mnras, 443,
  755

\bibitem[{{Husemann} {et~al.}(2011){Husemann}, {Wisotzki}, {Jahnke}, \&
  {S{\'a}nchez}}]{husemann2011}
{Husemann}, B., {Wisotzki}, L., {Jahnke}, K., \& {S{\'a}nchez}, S.~F. 2011,
  \aap, 535, A72

\bibitem[{{Husemann} {et~al.}(2008){Husemann}, {Wisotzki}, {S{\'a}nchez}, \&
  {Jahnke}}]{Husemann2008}
{Husemann}, B., {Wisotzki}, L., {S{\'a}nchez}, S.~F., \& {Jahnke}, K. 2008,
  \aap, 488, 145

\bibitem[{{Jarvis}(1990)}]{Jarvis1990}
{Jarvis}, B.~J. 1990, \aap, 240, L8

\bibitem[{{Kauffmann} {et~al.}(2003){Kauffmann}, {Heckman}, {Tremonti},
  {Brinchmann}, {Charlot}, {White}, {Ridgway}, {Brinkmann}, {Fukugita}, {Hall},
  {Ivezi{\'c}}, {Richards}, \& {Schneider}}]{kauffmann}
{Kauffmann}, G., {Heckman}, T.~M., {Tremonti}, C., {et~al.} 2003, \mnras, 346,
  1055

\bibitem[{{Kewley} {et~al.}(2001){Kewley}, {Dopita}, {Sutherland}, {Heisler},
  \& {Trevena}}]{Kewley2001}
{Kewley}, L.~J., {Dopita}, M.~A., {Sutherland}, R.~S., {Heisler}, C.~A., \&
  {Trevena}, J. 2001, \apj, 556, 121

\bibitem[{{Kewley} {et~al.}(2006){Kewley}, {Groves}, {Kauffmann}, \&
  {Heckman}}]{Kewley2006}
{Kewley}, L.~J., {Groves}, B., {Kauffmann}, G., \& {Heckman}, T. 2006, \mnras,
  372, 961

\bibitem[{{Kinney} {et~al.}(2000){Kinney}, {Schmitt}, {Clarke}, {Pringle},
  {Ulvestad}, \& {Antonucci}}]{Kinney2000}
{Kinney}, A.~L., {Schmitt}, H.~R., {Clarke}, C.~J., {et~al.} 2000, \apj, 537,
  152

\bibitem[{{Kr{\"u}hler} {et~al.}(2017){Kr{\"u}hler}, {Kuncarayakti}, {Schady},
  {Anderson}, {Galbany}, \& {Gensior}}]{Kruehler2017}
{Kr{\"u}hler}, T., {Kuncarayakti}, H., {Schady}, P., {et~al.} 2017, ArXiv
  e-prints, arXiv:1702.05430

\bibitem[{{Lelievre} {et~al.}(1984){Lelievre}, {Nieto}, {Horville}, {Renard},
  \& {Servan}}]{lalievre1984}
{Lelievre}, G., {Nieto}, J.-L., {Horville}, D., {Renard}, L., \& {Servan}, B.
  1984, \aap, 138, 49

\bibitem[{{L{\'o}pez-Cob{\'a}} {et~al.}(2017){L{\'o}pez-Cob{\'a}},
  {S{\'a}nchez}, {Moiseev}, {Oparin}, {Bitsakis}, {Cruz-Gonz{\'a}lez},
  {Morisset}, {Galbany}, {Bland-Hawthorn}, {Roth}, {Dettmar}, {Bomans},
  {Gonz{\'a}lez Delgado}, {Cano-D{\'{\i}}az}, {Marino}, {Kehrig}, {Monreal
  Ibero}, \& {Abril-Melgarejo}}]{carlos2016}
{L{\'o}pez-Cob{\'a}}, C., {S{\'a}nchez}, S.~F., {Moiseev}, A.~V., {et~al.}
  2017, \mnras, 467, 4951

\bibitem[{{Miley} {et~al.}(1981){Miley}, {Heckman}, {Butcher}, \& {van
  Breugel}}]{Miley1981}
{Miley}, G.~K., {Heckman}, T.~M., {Butcher}, H.~R., \& {van Breugel}, W.~J.~M.
  1981, \apjl, 247, L5

\bibitem[{{Osterbrock}(1989)}]{osterbrock}
{Osterbrock}, D.~E. 1989, {Astrophysics of gaseous nebulae and active galactic
  nuclei}

\bibitem[{{Perlman} {et~al.}(1999){Perlman}, {Biretta}, {Zhou}, {Sparks}, \&
  {Macchetto}}]{Perlman1999}
{Perlman}, E.~S., {Biretta}, J.~A., {Zhou}, F., {Sparks}, W.~B., \&
  {Macchetto}, F.~D. 1999, \aj, 117, 2185

\bibitem[{{S{\'a}nchez} {et~al.}(2004){S{\'a}nchez}, {Garcia-Lorenzo},
  {Mediavilla}, {Gonz{\'a}lez-Serrano}, \& {Christensen}}]{Sanchez2004}
{S{\'a}nchez}, S.~F., {Garcia-Lorenzo}, B., {Mediavilla}, E.,
  {Gonz{\'a}lez-Serrano}, J.~I., \& {Christensen}, L. 2004, \apj, 615, 156

\bibitem[{{S{\'a}nchez} {et~al.}(2015){S{\'a}nchez}, {P{\'e}rez},
  {Rosales-Ortega}, {Miralles-Caballero}, {L{\'o}pez-S{\'a}nchez},
  {Iglesias-P{\'a}ramo}, {Marino}, {S{\'a}nchez-Menguiano},
  {Garc{\'{\i}}a-Benito}, {Mast}, {Mendoza}, {Papaderos}, {Ellis}, {Galbany},
  {Kehrig}, {Monreal-Ibero}, {Gonz{\'a}lez Delgado}, {Moll{\'a}}, {Ziegler},
  {de Lorenzo-C{\'a}ceres}, {Mendez-Abreu}, {Bland-Hawthorn}, {Bekerait{\.e}},
  {Roth}, {Pasquali}, {D{\'{\i}}az}, {Bomans}, {van de Ven}, \&
  {Wisotzki}}]{sanchez2015a}
{S{\'a}nchez}, S.~F., {P{\'e}rez}, E., {Rosales-Ortega}, F.~F., {et~al.} 2015,
  \aap, 574, A47

\bibitem[{{S{\'a}nchez} {et~al.}(2016){S{\'a}nchez}, {P{\'e}rez},
  {S{\'a}nchez-Bl{\'a}zquez}, {Gonz{\'a}lez}, {Ros{\'a}lez-Ortega},
  {Cano-D{\'{\i}} az}, {L{\'o}pez-Cob{\'a}}, {Marino}, {Gil de Paz},
  {Moll{\'a}}, {L{\'o}pez-S{\'a}nchez}, {Ascasibar}, \&
  {Barrera-Ballesteros}}]{PIPE3D}
{S{\'a}nchez}, S.~F., {P{\'e}rez}, E., {S{\'a}nchez-Bl{\'a}zquez}, P., {et~al.}
  2016, \rmxaa, 52, 21

\bibitem[{{Sanders} {et~al.}(2003){Sanders}, {Mazzarella}, {Kim}, {Surace}, \&
  {Soifer}}]{Sanders}
{Sanders}, D.~B., {Mazzarella}, J.~M., {Kim}, D.-C., {Surace}, J.~A., \&
  {Soifer}, B.~T. 2003, \aj, 126, 1607

\bibitem[{{Schmitt} {et~al.}(2006){Schmitt}, {Calzetti}, {Armus}, {Giavalisco},
  {Heckman}, {Kennicutt}, {Leitherer}, \& {Meurer}}]{Schmitt2006}
{Schmitt}, H.~R., {Calzetti}, D., {Armus}, L., {et~al.} 2006, \apjs, 164, 52

\bibitem[{{Sol{\'o}rzano-I{\~n}arrea}
  {et~al.}(2001){Sol{\'o}rzano-I{\~n}arrea}, {Tadhunter}, \& {Axon}}]{Axon2001}
{Sol{\'o}rzano-I{\~n}arrea}, C., {Tadhunter}, C.~N., \& {Axon}, D.~J. 2001,
  \mnras, 323, 965

\bibitem[{{Sutherland} {et~al.}(1993){Sutherland}, {Bicknell}, \&
  {Dopita}}]{sutherland1993}
{Sutherland}, R.~S., {Bicknell}, G.~V., \& {Dopita}, M.~A. 1993, \apj, 414, 510

\bibitem[{{Tarenghi}(1981)}]{1981ESASP.162..145T}
{Tarenghi}, M. 1981, in ESA Special Publication, Vol. 162, Optical Jets in
  Galaxies, ed. B.~{Battrick} \& J.~{Mort}

\bibitem[{{Tilak} {et~al.}(2005){Tilak}, {O'Dea}, {Tadhunter}, {Wills},
  {Morganti}, {Baum}, {Koekemoer}, \& {Dallacasa}}]{tilak2005}
{Tilak}, A., {O'Dea}, C.~P., {Tadhunter}, C., {et~al.} 2005, \aj, 130, 2513

\end{thebibliography}
\end{document}